\documentclass[11pt]{article}
\usepackage{geometry}                % See geometry.pdf to learn the options
\usepackage{amsthm}
\usepackage{amssymb,amsmath}
\usepackage{xcolor}
\usepackage{hyperref}
\usepackage{bbm}

\newcommand{\E}{{\mathbb E}}
\newcommand{\R}{{\mathbb R}}

\newtheorem{thm}{Theorem}

\newtheorem{corollary}[thm]{Corollary}

\theoremstyle{remark}
\newtheorem{remark}{Remark}

\usepackage[style=authoryear,backend=bibtex,natbib=true]{biblatex}
\renewbibmacro{in:}{}
\begin{document}
  \title{Optimal Turnover, Liquidity, and Autocorrelation}

  \author{Bastien Baldacci\footnote{Quantitative Advisory Solutions, bastien.baldacci.qas@protonmail.com}, Jerome Benveniste\footnote{Courant Institute, NYU Tandon School, Ritter Alpha, ejb14@nyu.edu} and Gordon Ritter\footnote{Courant Institute, NYU Tandon School, Columbia University, Baruch College, and Ritter Alpha, \mbox{ritter@post.harvard.edu} }}

  \date{\today}
  
  \maketitle

\subsection*{Introduction}

One of the central problems faced by institutional investment managers is the proper management of trading costs.
For very active strategies, the most significant source of trading costs is usually market impact; see among others \citet{almgren2005direct,bouchaud2009price}.
The manager may reduce overall market impact cost by reducing turnover, but this potentially also 
reduces the manager's ability to monetize time-sensitive trading opportunities.

\medskip
The proper level of turnover of a real strategy can depend on many factors.  Nevertheless, in a class of Gaussian process models with linear price impact\footnote{This class of models has been the subject of an intensive literature, see among others \citet{almgren2001optimal,garleanu2016dynamic} and the reference books \citet{cartea2014buy,gueant2016financial}}, we derive a simple  
explicit relation \eqref{eq:optimal-turn-explicit} between the optimal turnover, the autocorrelation of the trading signal, the investor's risk-aversion, and the liquidity and volatility of the underlying asset. 

\medskip
Our explicit results, while mathematically elegant, only apply to the case of linear price impact, leading to quadratic total cost.
Empirical studies support the conclusion that the price impact of large orders
is in fact proportional to the square root of the order's participation;  see \citet{toth2011anomalous} and references therein.  Nevertheless, we join \citet{garleanu2013dynamic} in the belief that it is worthwhile to 
develop explicit formulas which apply to the linear case,
because they can be used as heuristics, approximations to more complex models, or to develop bounds.

\medskip
Historically, one of the first explicit formulas relating approximate investment performance to forecast accuracy was the \citet{grinold1989fundamental} fundamental law of active management, which asserts that $\mathrm{IR} \approx \mathrm{IC} \sqrt{N}$, where IR is the information ratio, IC is the information coefficient defined as the correlation of a single signal to the dependent return, and $N$ is the effective number of independent bets. Bets may be independent either because they pertain to statistically independent investments, or because they pertain to different periods of time, or both.

\medskip
Unfortunately the \citet{grinold1989fundamental} formula cannot relate the turnover of the strategy to any other meaningful quantity. The formula always prefers increasing turnover, because if asset returns are serially independent,
then trading more often increases $N$ with no perceived cost.
Liquidity of the asset is not an input to the formula. 

\medskip
In order to better understand the fundamental relation between optimal turnover, liquidity, and autocorrelation of alpha signals, we work in a continuous-time stochastic process model. Letting $x_t$ denote an investor's holdings of a
risky asset at time $t$, denominated in dollars (or any convenient numeraire), define the \emph{steady-state turnover} as the following limit
\begin{equation}\label{def:steady-state-turnover}
    \lim_{t \to \infty} \frac{\E|\dot x_t|}{\E |x_t|}
\end{equation}
where $\dot x_t$ denotes the time-derivative. 

\medskip
The steady-state turnover of the \emph{optimal} strategy is of clear interest  to practitioners and portfolio managers, as is the steady-state Sharpe ratio. 
In what follows, we show that in a convenient Gaussian process model, the steady-state turnover can be computed explicitly, and obeys a clear relation to the liquidity of the asset and to the autocorrelation of the alpha forecast signals. Thus, we contribute to the literature on optimal execution, in the spirit of \citet{cartea2016incorporating,lehalle2019incorporating}, by providing closed-form expressions of important trading metrics in a general linear-quadratic framework.
Let $\lambda$ denote the linear price impact coefficient in the tradition of  \citet{kyle1985continuous}, let $\kappa$ denote the investor's absolute risk-aversion, and let $\sigma$ denote the instantaneous volatility. In the special case of an Ornstein-Uhlenbeck model with mean-reversion speed $\phi$, we find that steady-state optimal turnover is given by 
\[
\sqrt{\frac{\sigma  \left(\phi  \sqrt{\kappa  \lambda }+\kappa  \sigma \right)}{\lambda}} .
\]
Before deriving these explicit formulas, we present Theorem \ref{thm:general-quadratic}, 
which is arguably the most general result on 
optimal trading strategies for quadratic trading-cost models in the setting of a 
mean-quadratic-variation objective.

\subsection*{A very general result on quadratic costs}

Presently we generalize the well-known result of \citet{garleanu2016dynamic}, 
which assumed the return-predictor process is a Markovian jump diffusion, 
to any square-integrable process. 
Our theorem also generalizes the main result of \citet{almgren2001optimal}, and thus
has applications to optimal execution of algorithmic orders.
Such a general result is of intrinsic interest in its own right, but we need it specifically
later in this paper, to derive results on steady-state optimal turnover and 
strategy performance. The proof is also of interest, as it shows that the 
problem can be converted into a convex optimization problem in an infinite-dimensional
space, and is thus amenable to standard convex optimization techniques as per 
\citet{baldacci2020note}. 

\medskip
Fix a probability measure space $(\Xi, \mathbb{P})$
where $\mathbb{P}$ is a probability measure on a $\sigma$-algebra of events. 
The outcomes in this space are various possible trajectories of the market and of 
our trading within it. 
The filtration $\mathcal{F} = \{ \mathcal{F}_t : t \geq 0\}$ 
denotes, as usual, 
the information that is available at time $t$. An \emph{adapted process} 
is a stochastic process $x$ such that each $x_t$ is $\mathcal{F}_t$-measurable.
Since it is used often, we define a notation for the 
conditional expectation: 
\[
    \E_t[Y] := \E[ Y \mid \mathcal{F}_t] ,
\]
where $Y$ is any random variable. 

\medskip
Let $\mathcal{H}$
denote the usual real Hilbert space of all adapted $\mathbb{R}^N$-valued processes on 
$[0, \infty)$ which are integrable in the mean-square sense: 
\[
\mathbb{E}\int_0 ^\infty ||\mu_t||^2 dt < \infty .
\]
Also let $\mathcal{A}$ be the Sobolev space of $\R^N$-valued adapted processes 
that are almost surely differentiable and whose derivative lies in $\mathcal{H}$. 
We use the symbol $\cdot$ for the Euclidean inner product in $\R^N$, and 
$\langle \, , \, \rangle$ for the standard inner product on $\mathcal{H}$: 
\[
    \langle x , y \rangle = \mathbb{E} \int_0 ^\infty x_t \cdot y_t \, dt \, .
\]

In the following, the positive-definite matrix $\Omega$ denotes the quadratic co-variation of the asset return process; with this interpretation, the risk term in our objective function is integrated instantaneous variance. This corresponds with the treatment of integrated variance in \citet{almgren2001optimal}, and also corresponds directly to the objective function of \citet{garleanu2016dynamic} in the zero-discounting case. 

\noindent
\begin{minipage}{\textwidth}
\begin{thm} \label{thm:general-quadratic}
Let $\mu \in \mathcal{H}$, 
let $\Omega$ be a positive-definite $N \times N$ covariance matrix and $\Lambda$ a positive-definite $N \times N$ market-impact coefficient matrix. Let $\kappa > 0$ be a risk-aversion coefficient. 
Then for any $x_0 \in \R^N$, there is a unique solution to the optimization problem
\begin{equation} \label{eq:general-quadratic-max}
    \max_{x \in \mathcal{A}; \; x(0) = x_0} 
    \E \int_0 ^\infty \left[ \mu_t \cdot x_t - \frac{1}{2} \dot x \cdot \Lambda \dot x - \frac{\kappa}{2}  x_t \cdot \Omega x_t \right] \, dt,
\end{equation}
given by the solution, with boundary condition $x(0) = x_0$, to the stochastically-forced ODE system 
\begin{equation} \label{eq:stoch-force-ode}
    \dot x_t = - \Gamma x_t + b_t
\end{equation}
where
\begin{align}\label{eq:capital-gamma}
    \Gamma &= (\kappa \Lambda^{-1} \Omega)^{1/2} \text{ and } 
    \\
    b_t &= \int_t ^\infty e^{\Gamma(t-s)} \Lambda^{-1} \mathbb{E}_t \mu_s \, ds . 
    \label{eq:defn-bt}
\end{align}
\end{thm}
\end{minipage}

\begin{remark}
The objective in \eqref{eq:general-quadratic-max} may be unbounded if 
    \[
    \mathbb{E}\int_0 ^\infty ||\mu_s||^2\, ds = \infty
    \]
as will be the case in many examples of interest, including stationary processes. However, suppose that $\mu$ satisfies the weaker condition
    \[
    \mathbb{E}\int_t ^\infty e^{-\epsilon s} ||\mathbb{E}_t \mu_s||^2\, ds < \infty
    \]
for any $t$ and any $\epsilon > \Gamma$. This condition ensures that $b_t$ is well-defined for all $t\geq 0$, and is satisfied by most realistic forecasts; for instance, those with at most linear growth. Then the truncated process $\mu_s^K = \mathbbm{1}_{\{s < K\}} \mu_s$ is in $\mathcal{H}$, hence Theorem \ref{thm:general-quadratic} gives an optimal strategy $x^K_s$ corresponding to $\mu_s^K$. Moreover, one may prove that on every bounded interval 
$[0,T]$, $x^K \to x$ as $K \to \infty$ where $x$ is the solution to the ODE \eqref{eq:stoch-force-ode} for the forecast process $\mu$, and convergence is in the mean-square sense on $[0,T]$ for the process and its first derivative.\footnote{A proof of this fact is available from the authors on request.} 
Thus an investor with an infinite horizon and no discounting will arrive at a well-defined investment strategy by using \eqref{eq:stoch-force-ode}, or equivalently by considering finite-horizon problems and letting the horizon tend to infinity. 
\end{remark}

\begin{remark}
The system \eqref{eq:stoch-force-ode} is said to be stochastically-forced because, when 
viewed from time $t=0$, $b_t$ is a stochastic process due to the presence of 
$\E_t\mu_s$, which is a conditional expectation in the filtration. 
\end{remark}

\begin{remark}
Note that the solution \eqref{eq:stoch-force-ode} can be rewritten so that we only take the square root of symmetric and positive semi-definite matrices using the fact that $$\Gamma = \kappa\Lambda^{-1/2}\big(\Lambda^{-1/2}\Omega\Lambda^{-1/2}\big)^{1/2}\Lambda^{1/2}.$$
\end{remark}

\medskip
\noindent \emph{Proof of Theorem \ref{thm:general-quadratic}.} Let $K$ be the integral operator defined by
\[
    (Ku)_t = \int_0 ^t u_s \, ds,
\]
which allows us to write the original problem as a convex 
optimization problem: 
\begin{equation} \label{eq:convex-analysis}
    \max_u \big[ \langle Ku, \mu \rangle - \frac{1}{2} \langle u, \Lambda u \rangle - \frac{\kappa}{2} \langle Ku, \Omega Ku \rangle
    \big] \, .
\end{equation}
Here $u$ is an auxiliary process defined so that $x_t = (Ku)_t$. 
By convex analysis, any solution to \eqref{eq:convex-analysis} must also solve
\begin{equation}
\label{eq:FOC0}
    K^* \mu - \Lambda u -\kappa K^* \Omega K u = 0.
\end{equation}
where the operator $K^*$ is given by 
\[
    K^* u = \int_t ^\infty \mathbb{E}_t u_s ds
\]
and so \eqref{eq:FOC0} takes the form
\begin{equation}
\label{eq:FOC1}
    \int_t ^\infty \E_t \mu_s \, ds - \Lambda u_t - \kappa \int_t ^{\infty} \int_0 ^s \Omega \E_t u_z \, dz ds  = 0  \, .
\end{equation}
Fix a time $t_0$, let $u$ be the solution of \eqref{eq:FOC0}, let $x = Ku$, and let
\[
    x^{t_0}_t = \mathbb{E}_{t_0} x_t, \; \text{ for } t > t_0.
\]
Applying the operator $\mathbb{E}_{t_0}$ to \eqref{eq:FOC1} and differentiating 
with respect to $t$ gives the following linear second-order ODE for 
$x^{t_0}_t, \; t \geq t_0$:
\begin{equation}
    \frac{d^2 x^{t_0}_t}{dt^2} =  \kappa \Lambda^{-1} \Omega x^{t_0}_t - \Lambda^{-1} \mathbb{E}_{t_0} \mu_t.
\end{equation}
Solving a linear second-order ODE is standard, and in this case yields: 
\begin{equation}
    x^{t_0}_t = e^{-\Gamma (t-t_0)} x^{t_0}_{t_0} + \int_{t_0} ^{t} e^{-\Gamma(t-s)}\int_{s} ^\infty e^{-\Gamma(z - s)}\Lambda^{-1} \mathbb{E}_{t_0} \mu_z \, dz ds
\end{equation}
and therefore
\[
    \left.\frac{dx^{t_0}_t}{dt}\right|_{t=t_0} = -\Gamma x_{t_0} ^{t_0} + b_{t_0}.
\]
Recalling that $x^{t_0}_{t_0} = x_{t_0}$, that 
\[
    \left.\frac{dx^{t_0}_t}{dt}\right|_{t=t_0} = \dot x_{t_0},
\]
and that $t_0$ was arbitrary gives the desired result. $\Box$

\medskip
This result is helpful due to its level of generality. Theorem \ref{thm:general-quadratic} represents a strict generalization of \citet{garleanu2016dynamic} Proposition 1. 
We generalize this result to predictor processes which are not necessarily Markov, and to the case of zero discounting. For Markov predictors with zero discounting, our solution agrees with \citet{garleanu2016dynamic} exactly. The rate matrix presented in \citet{garleanu2016dynamic} simplifies to
\[
    {\bar{M}}^{\text{rate}} = (\kappa \Lambda^{-1} \Omega)^{1/2} = \Gamma .
\]
Under these conditions, the matrix denoted $b$ in \citet{garleanu2016dynamic} is also equal to the rate matrix.
We then have from \eqref{eq:defn-bt}, 
\begin{align*}
    {\bar{M}}^{\text{aim}} &= \Gamma^{-1} \Lambda^{-1} \int_t ^\infty e^{\Gamma(t-s)} \mathbb{E}_t \mu_s \, ds
    \\
    &= \Gamma \int_t ^\infty e^{\Gamma(t-s)} \mathbb{E}_t [(\kappa \Omega)^{-1}\mu_s] \, ds
\end{align*}
The latter expression agrees with equation (12) of \citet{garleanu2016dynamic}. 

\medskip
Theorem \ref{thm:general-quadratic} 
enables researchers to derive explicit 
optimal trading strategies for a wide range of models by relating them to equivalent ODE systems. Moreover, the maximization in \eqref{eq:general-quadratic-max} is not simply over static, predetermined trading 
plans, but over the substantially larger class of all admissible stochastic
processes. This contributes to the literature on optimal execution in a linear-quadratic framework in the spirit of \citet{lehalle2019incorporating}. In our case, Theorem \ref{thm:general-quadratic} facilitates the derivation of
simple explicit formulas for the steady-state turnover and information ratio of the 
optimal strategy, as we shall present in the coming sections.

\subsection*{Steady-state calculations for Gaussian processes}

In the sequel we are interested in detailed calculations for single-asset trading 
paths and in particular, 
on account of \eqref{eq:stoch-force-ode}, we are interested in 
studying a Gaussian process $x$ defined by the stochastically-forced linear ODE: 
  \[
    \dot x_t = - \gamma x_t + a_t,
  \]
for some constant $\gamma > 0$, where $a_t$ is a process satisfying 
\begin{equation}\label{expdecay}
    \E_t a_s = e^{-\theta (t-s)} a_t \text{ for all }  s > t,
\end{equation}
for some $\theta > 0$. We shall see later that certain of the processes 
\eqref{eq:defn-bt} from the theorem must necessarily satisfy the relation 
\eqref{expdecay}, and hence are valid choices for $a_t$.
It follows from \eqref{expdecay} that $\lim_{s \to \infty} \E_t[a_s] = 0$, 
so the steady-state mean of the process $a$ vanishes, and hence the same must hold for $x$ and $\dot x$. For simplicity let us assume that the unconditional mean of $a_t$ is zero for all $t$.

\medskip
Under these assumptions, the Gaussian process $x_t$ has certain steady-state properties which can be derived in closed form. Since $x_t$ is prototypical solution of an optimal trading problem that we discuss later on, the steady-state properties of $x_t$ are clearly of crucial importance to understanding the optimal steady-state turnover. 

\medskip
For notational convenience, let us define the following three functions: 
\begin{equation}\label{eq:def-three-functions}
    h(t) = \E a_t x_t, 
    \  
    g(t) = \E x^2_t, 
    \text{ and }
    v(t) = \E \dot x^2_t . 
\end{equation}
We are interested in the steady-state behaviour of these functions. 

Note that $h$ has time-derivative 
  \begin{align*}
    \dot h &= \E[\dot a_t x_t + a_t \dot x_t]
    \\
    &= -\theta \E a_t x_t + \E a_t^2 - \gamma \E a_t x_t
    \\
    &= c_0 -(\theta + \gamma) h.
  \end{align*}
where $ c_0 := \E a_t^2$ denotes the steady-state variance of $a_t$. 
Then the steady-state value of $h(t)$ is
\[
  \bar h = \frac{c_0}{\theta + \gamma}.
\]
Similarly, 
\begin{align*}
  \dot g &= 2\E \dot x_t x_t = 2\E a_t x_t - 2 \gamma \E x_t^2\\
  &= 2h - 2 \gamma g,
\end{align*}
in steady state, that is
\[
  \dot g = 2 \bar h - 2 \gamma g,
\]
so the steady-state value for $g$ is
\[
  \bar g = \frac{\bar h}{\gamma}.
\]
Finally, we have
\begin{align*}
  v(t) &= \E[(a_t - \gamma x_t)^2]
  \\
  &= \E[ a^2_t - 2\gamma a_t x_t + \gamma^2 x_t^2]
  \\
  &= c_0 - 2\gamma h(t) + \gamma^2 g(t),
\end{align*}
which, in steady state is:
\begin{align*}
  \bar v &= c_0 - 2\gamma \bar h + \gamma^2 \bar g
  = c_0 - \gamma \bar h
  \\
  &= c_0 \left(1 - \frac{\gamma}{\theta + \gamma}\right)
  \\
  &= \frac{c_0 \theta}{\theta + \gamma} = \bar{h} \theta . 
 \end{align*}

Since differentiation is a linear operator, the derivative of a Gaussian process
is another Gaussian process (see \citet{williams2006gaussian} section 9.4 and references therein). 
Since $\{ x_t \}$ is a Gaussian process, each $x_t$ is a Gaussian random variable, as is each $\dot x_t$. In our example both $x_t$ and $\dot x_t$ have mean zero. From classical properties of the mean-zero normal distribution, one then has \mbox{$\E[ |x_t| ]^2 = (\pi / 2) \E[x_t^2]$}, and the same holds for $\dot x_t$ with the same constant, $\pi / 2$. 
Hence the steady-state turnover, as defined by 
\eqref{def:steady-state-turnover}, exists and is given by 
\begin{equation} \label{eq:main-turnover}
  \lim_{t \to \infty} \frac{\E|\dot x_t|}{\E |x_t|} 
  = 
  \frac{\sqrt{\bar v}}{\sqrt{\bar g}} = \sqrt{\theta \gamma}.
\end{equation}

\subsection*{Single-asset trading strategies}

Suppose an investor with risk-aversion $\kappa > 0$ is trading a single 
financial asset and solves
\begin{equation}
\label{mqvobj}
    \sup_x \E \left[ \int_0 ^\infty \mu_t x_t - \frac{1}{2} \kappa \sigma^2 x_t^2 
    - 
    \frac{1}{2} \lambda {\dot x_t}^2 \, dt \right]
\end{equation}
where $x, \mu$ are assumed to satisfy the technical hypotheses in 
Theorem \ref{thm:general-quadratic}, and the supremum is over all such $x$. 
The use of $\lambda$ to denote a linear price impact coefficient is inspired by, and consistent with, the notation of \citet{kyle1985continuous}; indeed such a coefficient is often referred to as \emph{Kyle's lambda}. 
We are interested in the steady-state properties of solutions to \eqref{mqvobj}, 
for which the key result is the following corollary of Theorem \ref{thm:general-quadratic}. 

\begin{corollary}\label{thm:folk} 
The optimal trading strategy solving (\ref{mqvobj}) is the solution of the ODE with stochastic coefficients:
\begin{equation}
\label{ODE}
    \dot x_t = - \gamma x_t + m_t,
\end{equation}
where:
\begin{align}
    \label{eq:def-gamma}
    \gamma &= \sqrt{\frac{\kappa \sigma^2} {\lambda}};
    \\
    \label{eq:def-m}
    m_t &= \int_t ^\infty \lambda ^{-1} e^{-\gamma(s-t)} \E_t \mu_s \, ds.
\end{align}
\end{corollary}
Corollary \ref{thm:folk} is a straightforward consequence of applying Theorem \ref{thm:general-quadratic}
to the single-asset case, where $N = 1$ and the $N \times N$ matrices 
$\Lambda, \Omega$ are just scalars $\lambda, \sigma^2$. In particular 
\eqref{eq:def-gamma} may be seen as the reduction of \eqref{eq:capital-gamma}. 

\medskip 
Assume now that the alpha-forecast process $\mu_t$ is 
an Ornstein-Uhlenbeck (O-U) process: 
\begin{equation} \label{eq:ou-process}
    d\mu_t = - \phi \mu_t dt + \nu\, dW.
\end{equation}
In this context $\phi$ is called the \emph{speed of mean-reversion}, 
and $\ln(2) / \phi$ is the half-life. 
Then $m_t$ in \eqref{eq:def-m} satisfies the exponential-decay 
condition \eqref{expdecay}, with $\theta = \gamma + \phi$:
\begin{equation}\label{eq:m-s-exponential-decay}
    \E_t m_s = e^{-(\gamma + \phi)(s - t) } m_t.
\end{equation}

Hence the calculations leading to \eqref{eq:main-turnover} apply. 
In particular, the optimal steady-state turnover exists and 
is given by \eqref{eq:main-turnover} which becomes: 
\begin{equation}\label{eq:optimal-turn-explicit}
  \text{optimal turnover} = \gamma \sqrt{\phi/\gamma + 1}.
\end{equation}
Plugging in \eqref{eq:def-gamma} it is easily verified that 
\eqref{eq:optimal-turn-explicit} is equal to $\sqrt{{\sigma  (\phi  \sqrt{\kappa  \lambda }+\kappa  \sigma )} / {\lambda}}$, and that 
$\gamma$ is the scalar version of the Garleanu-Pedersen rate matrix.

% ## Type the following commands into R to re-create this example 
% sigma = 0.01
% adv = 10*10^6
% bps = 10^(-4)
% lambda = 10 * bps / (0.01 * adv)
% kappa = 10^(-6)
% gamma = sqrt(kappa*sigma^2/lambda)
% phi = log(2)/5.0
% IR = 0.5 * sqrt(gamma/(2*phi*(phi+2*gamma)))

\medskip 
Eq.~\eqref{eq:optimal-turn-explicit}, while only strictly valid in the case 
of quadratic costs, can be used by practitioners as an easily-remembered rule of thumb.
One can see, e.g. from \eqref{eq:m-s-exponential-decay}, that $\gamma$ and $\phi$
each have dimensions of inverse time, and so optimal 
turnover is also in the same units as $\gamma$, inverse time. As such it represents
the fraction of steady-state book size which is traded in the given time unit. 
For example, if $\kappa = 10^{-6}$, then for an asset with $\sigma = 0.01/ \mbox{day}$, 
$\lambda = \mbox{10 bps per 1\% of ADV}$, and average daily volume (ADV) is 
$\$\,10\,\mbox{million}$, then 
one has $\gamma =0.1$/day. 
If $\phi = 0.2$/day, corresponding to a half-life about 3.5 days, 
then $\phi/\gamma = 2$ and optimal turnover 
\eqref{eq:optimal-turn-explicit} is about 
\mbox{$\gamma\sqrt{3} \approx$ 17.3\%/day}. 

From \eqref{eq:def-m} and the fact that
\[
\E_t \mu_s = e^{-\phi (s-t)} \mu_t,
\]
we get the relation
\begin{equation}\label{eq:m-versus-mu}
    m_t = \lambda^{-1} (\gamma + \phi)^{-1} \mu_t.
\end{equation}
Thus, using \eqref{eq:def-three-functions} with $m_t$ in place of $a_t$,
and also using \eqref{eq:m-versus-mu} to solve for $\mu_t$, we find that 
the expected \emph{ex ante} rate of profitability is 
\begin{equation}
    \E [x_t \mu_t]  = \lambda (\gamma + \phi ) \E[x_t m_t] = \lambda (\gamma + \phi ) h(t)
\end{equation}
We define the \emph{steady-state information ratio (IR)} as 
\[
    \text{IR} = \lim_{t \to \infty} 
        \frac{\E[x_t\mu_t - (\lambda/2) \dot x_t^2]}{\sqrt{ \E[ \sigma^2 x_t^2 ] }}
\]
The same calculations that were used in deriving 
\eqref{eq:main-turnover} now allow us to find, in the special case of the O-U process given in \eqref{eq:ou-process} for $d\mu_t$, that 
\begin{equation} \label{eq:steady-state-ir}
 \text{IR} = \frac{\nu}{2\sigma} 
    \sqrt{ \frac{\gamma}{2\phi(\phi + 2\gamma)} }
\end{equation}
In the numerical example from above, with $\phi/\gamma = 2$ one has 
has $\text{IR} =\nu / (8\gamma\sigma) \approx 0.5 \nu / \sigma$. 

\medskip 
The extension to a portfolio 
of $N$ \emph{statistically independent} assets, all governed by mean-reverting dynamics 
with similar half-lives and liquidity parameters, may be approximated 
by multiplying \eqref{eq:steady-state-ir} by 
$\sqrt{N}$ as per \citet{grinold1989fundamental}. In general the assets
in a portfolio will not be statistically independent, but there are special 
cases which may be approximated using the independence assumption. 
For example, in a market where 
the covariance matrix is well-described by a multi-factor model 
\citep{ross1976arbitrage}, 
then a strictly factor-neutral strategy may be modeled as trading synthetic 
``residual assets,'' which are defined as baskets that realize the factor model's residual
returns. For example, if one starts with a position in stock $i$, and then adds
an appropriate combination of pure factor portfolios to completely hedge stock $i$'s exposures 
to all common factors, one has a representation of the $i$-th residual asset as this hedged basket.

\medskip 
Residual assets will be statistically independent if the factor model's 
covariance matrix represents the true covariance of the data-generating process. 
For strategies modeled as dynamic portfolios of residual assets, the approximation 
of multiplying \eqref{eq:steady-state-ir} by $\sqrt{N}$ is reasonable. Thus one obtains 
the closed-form generalization of the classic ``fundamental law of active management''
to a modern context, in which the trader is aware of market impact and acts in a dynamically
optimal way with respect to a mean-quadratic-variation objective.

\printbibliography

\end{document}